\documentstyle[twoside]{article}

\catcode`\@=11
\long\def\@makefntext#1{
\protect\noindent \hbox to 3.2pt {\hskip-.9pt  
$^{{\eightrm\@thefnmark}}$\hfil}#1\hfill}		

\def\thefootnote{\fnsymbol{footnote}}
\def\@makefnmark{\hbox to 0pt{$^{\@thefnmark}$\hss}}	
	
\def\ps@myheadings{\let\@mkboth\@gobbletwo
\def\@oddhead{\hbox{}
\rightmark\hfil\eightrm\thepage}   
\def\@oddfoot{}\def\@evenhead{\eightrm\thepage\hfil
\leftmark\hbox{}}\def\@evenfoot{}
\def\sectionmark##1{}\def\subsectionmark##1{}}

\oddsidemargin=\evensidemargin
\renewcommand{\thefootnote}{\fnsymbol{footnote}}

\newcounter{sectionc}\newcounter{subsectionc}\newcounter{subsubsectionc}
\renewcommand{\section}[1] {\vspace{12pt}\addtocounter{sectionc}{1} 
\setcounter{subsectionc}{0}\setcounter{subsubsectionc}{0}\noindent 
	{\tenbf\thesectionc. #1}\par\vspace{5pt}}
\renewcommand{\subsection}[1] {\vspace{12pt}\addtocounter{subsectionc}{1} 
	\setcounter{subsubsectionc}{0}\noindent 
	{\bf\thesectionc.\thesubsectionc. {\kern1pt \bfit #1}}\par\vspace{5pt}}
\renewcommand{\subsubsection}[1] {\vspace{12pt}\addtocounter{subsubsectionc}{1}
	\noindent{\tenrm\thesectionc.\thesubsectionc.\thesubsubsectionc.
	{\kern1pt \tenit #1}}\par\vspace{5pt}}
\newcommand{\nonumsection}[1] {\vspace{12pt}\noindent{\tenbf #1}
	\par\vspace{5pt}}

\newcounter{appendixc}
\newcounter{subappendixc}[appendixc]
\newcounter{subsubappendixc}[subappendixc]
\renewcommand{\thesubappendixc}{\Alph{appendixc}.\arabic{subappendixc}}
\renewcommand{\thesubsubappendixc}
	{\Alph{appendixc}.\arabic{subappendixc}.\arabic{subsubappendixc}}

\renewcommand{\appendix}[1] {\vspace{12pt}
        \refstepcounter{appendixc}
        \setcounter{figure}{0}
        \setcounter{table}{0}
        \setcounter{lemma}{0}
        \setcounter{theorem}{0}
        \setcounter{corollary}{0}
        \setcounter{definition}{0}
        \setcounter{equation}{0}
        \renewcommand{\thefigure}{\Alph{appendixc}.\arabic{figure}}
        \renewcommand{\thetable}{\Alph{appendixc}.\arabic{table}}
        \renewcommand{\theappendixc}{\Alph{appendixc}}
        \renewcommand{\thelemma}{\Alph{appendixc}.\arabic{lemma}}
        \renewcommand{\thetheorem}{\Alph{appendixc}.\arabic{theorem}}
        \renewcommand{\thedefinition}{\Alph{appendixc}.\arabic{definition}}
        \renewcommand{\thecorollary}{\Alph{appendixc}.\arabic{corollary}}
        \renewcommand{\theequation}{\Alph{appendixc}.\arabic{equation}}
        \noindent{\tenbf Appendix \theappendixc #1}\par\vspace{5pt}}
\newcommand{\subappendix}[1] {\vspace{12pt}
        \refstepcounter{subappendixc}
        \noindent{\bf Appendix \thesubappendixc. {\kern1pt \bfit #1}}
	\par\vspace{5pt}}
\newcommand{\subsubappendix}[1] {\vspace{12pt}
        \refstepcounter{subsubappendixc}
        \noindent{\rm Appendix \thesubsubappendixc. {\kern1pt \tenit #1}}
	\par\vspace{5pt}}

\newcommand{\textlineskip}{\baselineskip=13pt}
\newcommand{\smalllineskip}{\baselineskip=10pt}

\def\eightcirc{
\begin{picture}(0,0)
\put(4.4,1.8){\circle{6.5}}
\end{picture}}
\def\eightcopyright{\eightcirc\kern2.7pt\hbox{\eightrm c}} 

\newcommand{\copyrightheading}[1]
	{\vspace*{-2.5cm}\smalllineskip{\flushleft
	{\footnotesize International Journal of Modern Physics A, #1}\\
	{\footnotesize $\eightcopyright$\, World Scientific Publishing
	 Company}\\
	 }}

\def\abstracts#1#2#3{{
	\centering{\begin{minipage}{4.5in}\baselineskip=10pt\footnotesize
	\parindent=0pt #1\par 
	\parindent=15pt #2\par
	\parindent=15pt #3
	\end{minipage}}\par}} 

\newcommand{\bibit}{\nineit}

\renewenvironment{thebibliography}[1]
	{\frenchspacing
	 \ninerm\baselineskip=11pt
	 \begin{list}{\arabic{enumi}.}
	{\usecounter{enumi}\setlength{\parsep}{0pt}
	 \setlength{\leftmargin 12.7pt}{\rightmargin 0pt} 
	 \setlength{\itemsep}{0pt} \settowidth
	{\labelwidth}{#1.}\sloppy}}{\end{list}}

\newcounter{itemlistc}
\newcounter{romanlistc}
\newcounter{alphlistc}
\newcounter{arabiclistc}

\newcommand{\fcaption}[1]{
        \refstepcounter{figure}
        \setbox\@tempboxa = \hbox{\footnotesize Fig.~\thefigure. #1}
        \ifdim \wd\@tempboxa > 5in
           {\begin{center}
        \parbox{5in}{\footnotesize\smalllineskip Fig.~\thefigure. #1}
            \end{center}}
        \else
             {\begin{center}
             {\footnotesize Fig.~\thefigure. #1}
              \end{center}}
        \fi}

\newcommand{\tcaption}[1]{
        \refstepcounter{table}
        \setbox\@tempboxa = \hbox{\footnotesize Table~\thetable. #1}
        \ifdim \wd\@tempboxa > 5in
           {\begin{center}
        \parbox{5in}{\footnotesize\smalllineskip Table~\thetable. #1}
            \end{center}}
        \else
             {\begin{center}
             {\footnotesize Table~\thetable. #1}
              \end{center}}
        \fi}

\def\@citex[#1]#2{\if@filesw\immediate\write\@auxout
	{\string\citation{#2}}\fi
\def\@citea{}\@cite{\@for\@citeb:=#2\do
	{\@citea\def\@citea{,}\@ifundefined
	{b@\@citeb}{{\bf ?}\@warning
	{Citation `\@citeb' on page \thepage \space undefined}}
	{\csname b@\@citeb\endcsname}}}{#1}}

\newif\if@cghi
\def\cite{\@cghitrue\@ifnextchar [{\@tempswatrue
	\@citex}{\@tempswafalse\@citex[]}}
\def\citelow{\@cghifalse\@ifnextchar [{\@tempswatrue
	\@citex}{\@tempswafalse\@citex[]}}
\def\@cite#1#2{{$\null^{#1}$\if@tempswa\typeout
	{IJCGA warning: optional citation argument 
	ignored: `#2'} \fi}}

\def\pmb#1{\setbox0=\hbox{#1}
	\kern-.025em\copy0\kern-\wd0
	\kern.05em\copy0\kern-\wd0
	\kern-.025em\raise.0433em\box0}

\def\fnt#1#2{\footnotetext{\kern-.3em
	{$^{\mbox{\scriptsize #1}}$}{#2}}}

\def\fpage#1{\begingroup
\voffset=.3in
\thispagestyle{empty}\begin{table}[b]\centerline{\footnotesize #1}
	\end{table}\endgroup}

\def\runninghead#1#2{\pagestyle{myheadings}
\markboth{{\protect\footnotesize\it{\quad #1}}\hfill}
{\hfill{\protect\footnotesize\it{#2\quad}}}}
\font\tenrm=cmr10
\font\tenit=cmti10 
\font\tenbf=cmbx10
\font\bfit=cmbxti10 at 10pt
\font\ninerm=cmr9
\font\nineit=cmti9

\font\eightrm=cmr8

\def\qed{\hbox{${\vcenter{\vbox{			
   \hrule height 0.4pt\hbox{\vrule width 0.4pt height 6pt
   \kern5pt\vrule width 0.4pt}\hrule height 0.4pt}}}$}}

\renewcommand{\thefootnote}{\fnsymbol{footnote}}	

\begin{document}

\runninghead{Proton Decay in SO(10) SUSY GUTs} {Proton Decay
in SO(10) SUSY GUTs }

\normalsize\textlineskip
\thispagestyle{empty}
\setcounter{page}{1}

\copyrightheading{}			

\vspace*{0.88truein}

\fpage{1}
\centerline{\bf PROTON DECAY IN SO(10) SUPERSYMMETRIC }
\vspace*{0.035truein}
\centerline{\bf GRAND UNIFIED THEORIES \footnote{This talk is based on
work done in collaboration with A. Mafi and S. Raby.}
\footnote{Ohio state university preprint OHSTPY--HEP--T--00--018. }}
\vspace*{0.37truein}
\centerline{\footnotesize RADOVAN DERM\' I\v SEK }
\vspace*{0.015truein}
\centerline{\footnotesize\it Department of Physics, The Ohio
State University, 174 West 18th Ave.,}
\baselineskip=10pt
\centerline{\footnotesize\it Columbus, OH 43210, USA }
\vspace*{0.225truein}

\vspace*{0.21truein}
\abstracts{We calculate the proton lifetime in an
SO(10) supersymmetric grand unified theory [SUSY GUT] with
U(2) family symmetry. This model fits the low
energy data, including the recent data for neutrino oscillations. We
discuss the predictions of this model for the proton lifetime
in light of recent SuperKamiokande results which significantly
constrain the SUSY parameter space of the model.}{}{}

\textlineskip			
\vspace*{12pt}			

\noindent
Some of the nicest features of SUSY GUTs are gauge and Yukawa coupling
unification and also the interesting prediction of proton decay. 
Interesting because this prediction has a chance to be observed or
excluded by SuperKamiokande and Soudan II experiments.
In SUSY GUTs with an additional
family symmetry, acting horizontally
between generations, the hierarchy between generations can be generated by
sequential spontaneous breaking of this symmetry. Several predictive
models of this type were constructed. 
In this talk we focus on the models with a
$U(2)$ or, its discrete subgroup, $D_3$ family symmetry.$^1$ These models
fit the low energy data quite well including the recent data for neutrino
oscillations. In what follows we calculate the proton lifetime in these
models: we review the origin of proton decay in SUSY GUTs, discuss
parameters which are crucial for the calculation of the proton lifetime
and
compare results with recent experimental limits. For a
more detailed discussion see Ref.$^2$ and references
therein.

The dominant contribution to proton decay in SUSY GUTs involves dimension
five
baryon and lepton number violating operators. These operators result from
the exchange of color triplet Higgsinos which below the GUT scale are
integrated out resulting in suppresion by one power of the effective color
triplet mass $\tilde M_t$. 
The effective four--fermion baryon and lepton number violating operators
are obtained by dressing these dimension five operators by gluino,
chargino and neutralino exchanges.

\pagebreak

\textheight=7.8truein   
\setcounter{footnote}{0}
\renewcommand{\thefootnote}{\alph{footnote}}

The dominant decay mode of the proton is $ p \; \rightarrow \; K^+ \, 
\bar \nu $ and the amplitude for this process can be written in the
following way:
\begin{equation}
T \, (\,  p  \rightarrow  K^+ \,   
 \bar \nu )\; \simeq \;  ( FF ) \; \frac{ M_{1/2} }{ m^2_0 }
\; \frac{1}{ \tilde M_t} \; \beta_{lat} \; ,
\end{equation} 
\noindent
where $( FF )$ is the flavor factor depending on Yukawa and  
gauge couplings ( it is fixed once we fit the low energy data ), $M_{1/2}
/ m^2_0$ is a good approximation to a 
gaugino--scalar--scalar loop factor for $M_{1/2} <<  m_0$ , $\tilde M_t$
is an effective color triplet mass and $\beta_{lat}$ is a chiral
Lagrangian factor. In what follows we discuss the allowed regions of these
parameters and use them to find the upper bound on the proton lifetime.
In this talk we focus on the large $\tan \beta$ regime ( discussion
concerning the low $\tan \beta$ regime can be found in Ref.$^2$ ).

The chiral Lagrangian factor $\beta_{lat}$ appears in the calculation of
the matrix element of the four--fermion operator between a proton and
lepton + meson final state. The recent lattice calculation gives
$\beta_{lat} \, = \, 0.015 \;  GeV^3$ with a very small statistical
uncertainty ($\pm 1$ in the last digit).$^3$
We use this central value in our calculation. However the systematic
uncertaities connected with the chiral Lagrangian approach and the
quenched approximation may be significant.

To suppress proton decay the effective color triplet mass $\tilde M_t$ has
to be very large. It is a free parameter constrained only by requiring
perturbative threshold corrections to gauge coupling unification.
If we define the GUT scale $M_G$ and the unified gauge coupling at the
point where $\alpha_1$ and $\alpha_2$ meet: $\alpha_G ( M_G ) = \alpha_1 (
M_G ) = \alpha_2 ( M_G )$ then we typically need a correction to
$\alpha_3$
at the GUT scale ( given by $\epsilon_3 = 
( \alpha_3 (M_G ) - \alpha_G ( M_G ) ) / \alpha_G ( M_G )$ )
about $- 2 \% $ to $- 4 \%$ to obtain $\alpha_3 ( M_Z ) = 0.119 $.
The one loop threshold correction from the Higgs sector ( $ 5 + \bar
5$ ) only is given by$^4$
\begin{equation}
\epsilon_3  \, ( \ Higgs \ )
\; = \; \frac{3 \, \alpha_G}{5 \, \pi}
\, \log \, (\,
\frac{ \tilde M_t }{M_G} \, ) \, .
\end{equation}
If the maximum allowed threshold correction from the rest of the GUT
sector is $ - 10 \% $ then $\epsilon_3  \, ( \ Higgs \ )$ can be at most $
6 \% $ which through Eq. (2) translates into the upper bound for the
effective color
triplet mass $ \tilde M_t = 8 \times 10^{19} \; GeV $.

As can be seen from Eq. (1) in order to suppress proton decay, gaugino
masses should be as small as possible.
To be consistent with present experimental
bounds on gaugino masses we take the universal gaugino mass at the GUT
scale
to be $M_{1/2} = 175 \; GeV$.  

On the other hand squarks and sleptons should be as heavy as possible.
However in order for SUSY to solve the gauge hierarchy problem squark and
slepton masses are expected to be near the weak scale. Otherwise fine
tuning is necessary to obtain $M_Z \sim m_{Higgs} << \Lambda_{SUSY}$. Thus
this naturalness criteria places an upper bound on squark and slepton
masses. In fact it affects mostly squarks and sleptons of the third
generation since the first two generations couple weakly to the Higgs
boson. Furthermore due to renormalization group running the third
generation scalars are naturally lighter than the scalars of the first
two generations. Thus if we demand that the scalars of the
third generations are lighter than $1 \; TeV$ we can take the universal
scalar mass at the GUT scale to be $m_0 = 3 \; TeV$. 

With all these ingredients we calculate the theoretical upper bound on the
proton lifetime:
\vspace*{13pt}
\begin{center}
$ \tau \, (\,  p  \rightarrow  K^+ \,
 \bar \nu )\; = \; 4.7 \times 10^{33} \; $ years.
\end{center}
\vspace*{13pt}
\noindent
This should be compared with the latest SuperKamiokande $90 \%$ CL bounds
on proton decay based on 61--ktonyear exposure:$^5$
\vspace*{13pt}
\begin{center}
$ \tau_{exp} \, (\,  p  \rightarrow  K^+ \,
 \bar \nu )\; > \; 1.9 \times 10^{33} \; $ years.
\end{center}
\vspace*{13pt}

From this comparison several general conclusions can be drawn. 
SuperKamio- kande bounds on the proton lifetime severely constrain
SO(10) SUSY
GUTs.
To stay
above the experimental limit on proton decay gaugino masses have to be
near the allowed experimental lower bounds while squark and slepton masses 
have to be near the upper bounds allowed by naturalness. Furthermore we
have to allow a $ 6 \%$ threshold correction at the GUT scale coming from
the Higgs  
sector. Thus the theoretical upper bound on the proton lifetime we present
is
to be considered as very conservative. Nevertheless it is barely
consistent
with the experimental limit.

\nonumsection{Acknowledgements}
\noindent
I would like to thank my collaborators A. Mafi and S. Raby and organizers
of the DPF2000. This work was supported in part by DOE grant
DOE/ER/01545-787.

\nonumsection{References}

\end{document}